\documentclass[12pt]{article}
\usepackage{graphicx,a4} 
\usepackage{amssymb}
\usepackage{latexsym}
\begin{document}

\newcommand{\preprintno}[1]
{\vspace{-2cm}{\normalsize\begin{flushright}#1\end{flushright}}\vspace{1cm}}

\title{\preprintno{{\bf preprint}}
Decay of Excited String States at Colliders and Cosmic Ray Detectors}
\author{Malcolm Fairbairn\thanks{malc@physto.se}\\
{\em Cosmology, Particle astrophysics and String theory}\\ 
{\em Department of Physics, Stockholm University}\\
{\em AlbaNova Centre, SE-106 91, Stockholm, Sweden}}

\date{September 2005}

\maketitle
\begin{abstract}
It is argued that in the context of TeV gravity with large extra dimensions, excited string states produced in colliders and in the interaction of cosmic rays with the atmosphere may decay preferentially into invisible bulk modes, rather than visible gauge fields on the brane.  This contrast to the black hole case comes about because of the absence of a relationship between physical size and temperature for string ball states.  We estimate the effect of this upon the number of events predicted at cosmic ray observatories and colliders.
\end{abstract}

\section{Introduction}
One of the successes of string theory is its ability to provide a rather convincing description of the end point of black hole evaporation \cite{widje}. It also provides a semi-complete description of the high energy scattering of strings, the creation of excited string states and eventually black holes (or at least, a mapping out of the states which are thought to inhabit that region of parameter space \cite{veneziano,polchinski,damour}).  Despite these advances, it was thought for many years that these high energy aspects of the theory would never be tested in a laboratory because of the large value of the Planck mass.  However, over the past decade more freedom has emerged in the relationship between the 4D Planck mass, the higher dimensional Planck mass and the string scale through the arrival of a set of theories with large extra dimensions.  In particular, the use of branes where gauge fields can be confined has lead to the suggestion that the higher dimensional string and Planck scales may not be too far above the reach of present day colliders \cite{ADD,antoniadis}.  While it is perhaps unlikely that this is actually the case, the possibility has lead to interesting research into the signatures of trans-Planckian collisions at colliders and cosmic ray observatories producing black holes \cite{giddings,luis} and excited string states \cite{stringballs,sarkar} which then decay rapidly via Hawking or Hagedorn radiation.

In such theories when a black hole is formed, it is typically much smaller than the radius if the large extra dimensions, and since the temperature of a black hole is simply its inverse radius (disregarding numerical factors) very many of the Kaluza Klein (KK) modes of the fields in the bulk are kinematically available for production.  There was some debate as to whether this would prevent the detection of the decay products of the Black holes, since they might be emitted in the form of bulk modes rather than the visible standard model modes residing on the brane.  The issue was resolved however in a paper by Emparan, Horowitz and Myers where it was pointed out that although there are indeed a large number of states in the bulk to excite, the geometrical overlap of the black hole with those states gives rise to a suppression factor which exactly cancels the effect of the large phase space \cite{EHM}.

In a weakly coupled string theory, the situation at centre of mass energies close to the string scale is not well described by higher dimensional general relativity.  As the centre of mass energy rises above the string scale, there are a series of phenomena which occur, starting with the production of long excited string states which might be viewed as a chain of connected string bits each with length equal to the string length $l_s$.  Since one end of each string bit will perform a random walk relative to the other, the size of these objects must be calculated statistically \cite{turok,damour}.  The result for the radius of the string ball is
\begin{equation}
r_{SB}=l_s\sqrt{M_{SB}l_s}.
\end{equation}
These string balls also emit quanta at the Hagedorn temperature, $T_H$ \cite{amatirusso} which is related to the string length by
\begin{equation}
T_H=\frac{1}{2\sqrt{2}\pi l_s}.
\end{equation}
As the centre of mass energy rises, the string ball spectrum is distorted by self gravity, and the most probable configuration shrinks down so that when the centre of mass energy is large enough to create a black hole with a Schwarzschild radius equal to the string length, the corresponding excited string configuration also has size equal to the string length \footnote{This is a gross oversimplification of the situation, many more details can be found in the paper by Damour and Veneziano \cite{damour}}.  Table \ref{stringtable} is an attempt to summarize the situation for different centre of mass energies $M$ and different string couplings $g$.

\begin{table*}
\begin{center}
\begin{tabular}{|c|c|c|c|}
\hline
Energy & Object & $\sigma(M)$&Most probable size \\ \hline \hline
&&&\\
$1<Ml_s<1/g$& excited string& $g^2M^2l_s^4$&  \\ 
&doing random&&$M^{1/2}l_s^{3/2}$  \\ \cline{1-1} \cline{3-3}
&walk&&\\
&& unitarity prevents&  \\ \cline{2-2} \cline{4-4}
$1/g<Ml_s<1/g^{2}$&self-gravity of&growth of cross-&\\ 
&excited string&section beyond $l_s^2$&$1/(g^2M)<r<M^{1/2}l_s^{3/2}$\\
& restricts growth& & \\ \hline
&&&\\
$1/g^2<Ml_s$ & black hole & $r_{BH}^2$& $g^{2/(D-3)} l_s(l_sM)^{1/(D-3)}$ \\ 
&&&\\ \hline
\end{tabular}
\caption[]{\label{stringtable}\it Different outcomes for different centre of mass scattering energies $M$ and string couplings $g$ in a $D$-dimensional space time.}
\end{center}
\end{table*}

The purpose of this letter is to argue that although black holes will not decay preferentially into the bulk, it is not clear if the same is true for string balls, and indeed we claim that as the mass of the string ball grows, emission of quanta into bulk modes is favoured over emission into excitations on the brane.  We will start by reviewing the argument for black holes, following closely reference \cite{EHM}.  
\section{Black Holes Decay mainly on the Brane}
The black body formula for thermal emission from a sphere at temperature $T$ radiating into D space-time dimensions is given by ($\hbar=c=1$)
\begin{equation}
\frac{dE_D}{dt}=\sigma_D S_{D-2}r^{D-2} T^{D}
\end{equation}
where the area of a n-sphere is $A_n=S_n r^{n}$ with $S_n$ given by
\begin{equation}
S_n=\frac{2\pi^{n/2}}{\Gamma(n/2)}
\end{equation}
I am using the convention that a normal sphere is a 2-sphere and a circle is a 1-sphere. The Boltzmann constant in $D$ dimensions takes the form
\begin{equation}
\sigma_D=\frac{S_{D-3}}{(2\pi)^{D-1}(D-2)}\Gamma(D)\zeta(D)
\end{equation}

The temperature of the Hawking radiation for a $D$ dimensional black hole is given by
\begin{equation}
T_{BH}=\frac{D-3}{4\pi r_{BH}}
\end{equation}
We will be interested in calculating the emission of Hawking radiation into the whole space and comparing it with the emission into our 3 decompactified dimensions.  We therefore write the expression for the luminosity of the $D$-dimensional black hole into $D'$ dimensions ($D'\le D$) which is given by
\begin{equation}
\frac{dE_D}{dt}=\frac{S_{D'-3}}{(2\pi)^{D'-1}(D'-2)}\Gamma(D')\zeta(D')S_{D'-2}\left(\frac{D-3}{4\pi}\right)^{D'} \frac{1}{r_{BH}^2}=K(D,D')\frac{1}{r_{BH}^2}
\label{luminosity}
\end{equation}
so that apart from the dimensionless factor K(D,D'), there is no strong dependence upon the number of dimensions as to how quickly the black hole emits energy.  This is basically the statement that an equal amount of radiation will be emitted into the KK tower of a single bulk species as will be emitted into a single brane species.

\section{Hagedorn rain falls mainly off the brane?}

Now we will try and convince the reader that the same is not true for string ball events.

The heuristic explanation for the result for black holes outlined above goes as follows - the temperature of the black hole $T_{BH}\sim r_{BH}^{-1}$ and the luminosity $\dot{E}\sim A_D T^{D}$ where the $D-$dimensional area $A_D\sim r_{BH}^{D-2}$ so that for any space-time dimension $D$, the luminosity into those dimensions $\dot{E}\sim r_{BH}^{-2}$. The same is not true for string balls, since the relationship between the temperature and the size of the ball is not the same as for the black hole case.

The luminosity of a string ball into $D$ dimensions can be written in the same way
\begin{equation}
\frac{dE_D}{dt}=\frac{S_{D-3}}{(2\pi)^{D-1}(D-2)}\Gamma(D)\zeta(D)S_{D-2}\left(\frac{1}{2\sqrt{2}\pi}\right)^D\frac {\left(l_s M\right)^{(D-2)/2}}{l_s^2} 
\label{lumst}
\end{equation}
so now apart from the dimensionless numerical factors we can see that the luminosity depends not only on the mass of the string ball, but also on the total number of space-time dimensions, breaking the symmetry between area and temperature.
Another way of looking at this is to go to four dimensions where the bulk KK modes are viewed as massive particles.  Again, following closely the reasoning of \cite{EHM} and dropping numerical factors, we can write the emission rate per unit frequency into modes of momenta ${\bf k}$
\begin{equation}
\frac{dE}{d\omega dt}(\omega,{\bf k })\simeq\left(\omega^2-m^2\right)\frac{\omega A_D}{e^{\beta\omega}-1}d^{D-4}k
\end{equation}
where $A_D=r^{D-4}A_4$ is the surface area of the emitting object in the $D-$dimensional bulk.  For a KK mode with mass $m\ll T$ we can set $d^{D-4}k=(1/R)^{D-4}$ where $R$ is the radius of the compact space then we get the expression for emission into a single mode in the bulk.
\begin{equation}
\frac{dE}{d\omega dt}(\omega,m)\simeq\left(\frac{r_{SB}}{R}\right)^{D-4}\left(\omega^2-m^2\right)\frac{\omega A_4}{e^{\beta\omega}-1}
\label{lum}
\end{equation}
This is the same result as for normal emission into a 4D mode but with an additional suppression factor which reflects the small overlap of the wave-function of the emitter and the modes in the bulk.  For a black hole, the number of modes in the bulk which are kinematically available provides a large phase space factor $(T_{BH}R)^{D-4}$ which exactly cancels the suppression factor.  Since the string ball has a radius $r_{SB}=l_s\sqrt{l_s M}$ its overlap with the $D-4$ dimensional bulk suppresses the emission into light bulk modes by the factor
\begin{equation}
\frac{\dot{E}_{single\, bulk\, mode}}{\dot{E}_{single\, brane\, mode}}=\left(\frac{r_{SB}}{R}\right)^{D-4}=\left[\frac{l_s}{R}\sqrt{l_s M}\right]^{D-4}
\label{sup}
\end{equation}
whereas since the temperature of the string ball is a constant, the number of light modes with mass $m_{KK}<T_H$ which become available is also a constant
\begin{equation}
\#_{light\, bulk\, modes}\simeq(T_H R)^{D-4}\simeq\left(\frac{R}{l_s}\right)^{D-4}
\label{numkk}
\end{equation}
So that multiplying the luminosity into a single bulk mode (\ref{lum}) by the number of available modes (\ref{numkk}) and integrating over $\omega$ we get (up to numerical factors) the same expression for the luminosity into the bulk as equation (\ref{lumst})
\begin{equation}
\frac{dE}{dt}\simeq A_4\left(\frac{r_{SB}}{R}\right)^{D-4}\left(\frac{R}{l_s}\right)^{D-4}\simeq\frac{r_{SB}^{D-2}}{l_s^{D-4}}=\frac{(l_sM)^{(D-2)/2}}{l_s^2}
\end{equation}
\begin{figure}[h]
\includegraphics[height=15cm,width=10cm,angle=270]{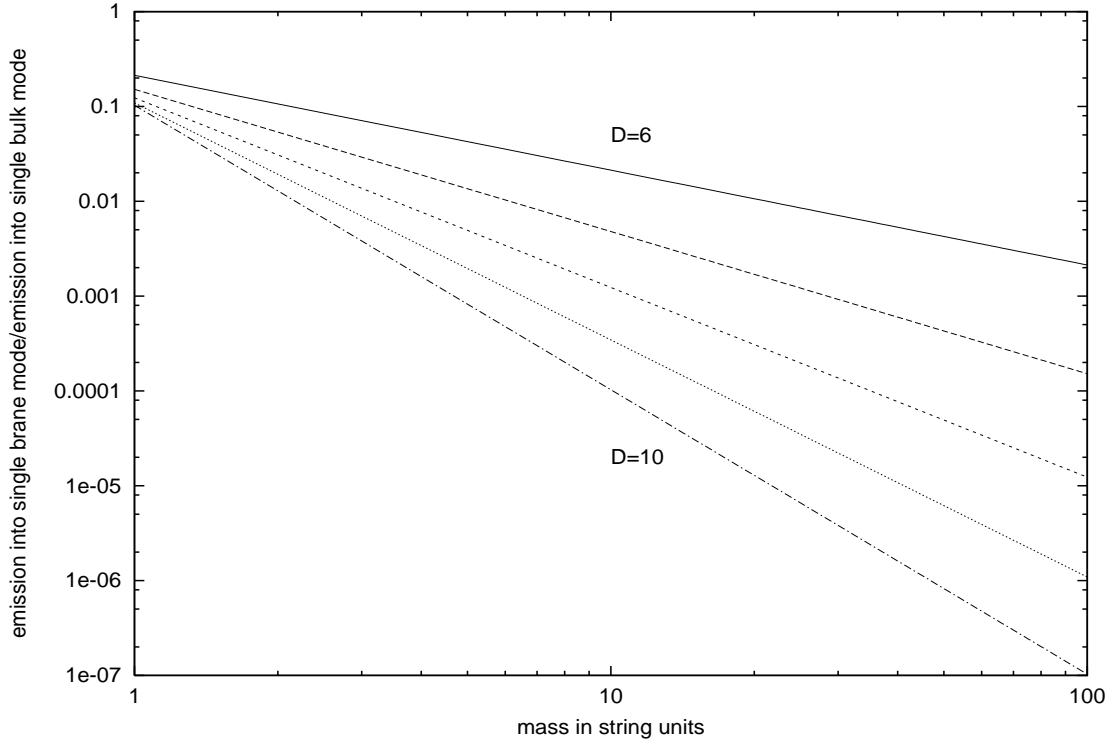} 
\caption{\it Ratio of luminosity of quanta emitted from a weakly coupled excited string ball into a single brane field vs.the KK tower of excitations of a single Bulk field. These estimates are only valid so long as $M\ll M_{BH}^{min}$ (see text).}
\label{ratio}
\end{figure}
Figure (\ref{ratio}) shows the ratio between the emission of energy onto a single brane mode vs. the emission into a single bulk species.  For large mass and a large number of dimensions, the string ball will emit preferentially into the bulk rather than the brane.  The figure is only valid for masses $M$ much less than the minimum mass for black hole formation $M_{BH}^{min}$ in order for one be sure that the extended string ball is the correct description of the excitation. $M_{BH}^{min}$ is given by the mass $M$ at which the Schwarzschild radius corresponding to $M$ is equal to the string length.
\begin{equation}
r_{BH}^{min}=g^{2/(D-3)}l_s(l_sM_{BH}^{min})^{1/(D-3)}=l_s\qquad \rightarrow\qquad M_{BH}^{min}=\frac{1}{g^2l_s}
\end{equation}
As the mass of the string ball approaches this limit, the probable size of the string ball either smoothly shrinks, or a second entropy minimum in the ensemble of configurations develops with a size closer to the string scale and then there would be something not completely unlike a tunneling phase transition to a smaller sized state.

\section{Discussion}

How interesting is this result, and what implications does it have for predictions of fluxes at colliders and cosmic ray detectors?  A first glance at the plots seem to suggest that the vast majority of emission will be into the bulk, however there are more than 100 degrees of freedom in the standard model which would correspond to brane modes, and far fewer in the bulk, namely the KK modes of the graviton assuming any other light scalars have been given masses. Most of these are the 72 quarks and 8 gluons of QCD which might lead to ambiguous signals at colliders due to jet fragmentation.  It is not clear that this ambiguity carries over into the situation at a cosmic ray detector where the air shower develops in an optically thick medium, erasing many details other than the initial energy exchange.

We have said that our results are only valid at masses well below $g^{-2}l_s^{-1}$ so the range over interesting suppression effects might occur is governed strongly by the string coupling.  Perhaps the simplest model of a TeV string scale is that of Antoniadis et al \cite{antoniadis} where the string coupling is of the same order as the fine structure constants $\alpha_i$ on the brane.  In this case, since the $\alpha_i$ on the brane should be between 0.1 and 0.01 we see there is quite a large regime where interesting suppression effect may occur.  If we look at figure \ref{ratio} we see that only if the string coupling is rather large so that the maximum mass of a string ball is around 10 $l_s^{-1}$ and the number of extra dimensions is small will the enhanced emission into the bulk be balanced by the large number of species on the brane.

It is expected that there is a flux of very high energy neutrinos which are produced during the interactions responsible for the GZK break in the cosmic ray spectrum \cite{spectrum}\cite{GZK}.  If the string scale is indeed around 1 TeV, these neutrinos should lead to the production of string balls and black holes when they hit nucleons in the atmosphere which subsequently decay and form showers.  The multiplicity of the decay is given roughly by the entropy of the object created, so for a string ball of mass $20 l_s^{-1}$ one would expect of the order of 20 emitted particles, however, if the reasoning made above is correct, many of these decays would be into the bulk rather than onto the brane, until the string ball shrinks down to a size closer to the string scale, and its decay onto the brane becomes dominant.

A characteristic signal for the suppression predicted in this paper at cosmic ray detectors would therefore be an increase in the number of events with reconstructed energy at the string scale, since objects with mass much larger than the string scale would radiate most of their energy into invisible bulk modes.  Then there would be a gap in the spectrum until the energy corresponding to the onset of black hole production, when decay onto the brane would resume \cite{EHM}.

At colliders, the situation is different because there is expected to be a large rate of black hole and/or string ball production if the string scale is in the TeV range \cite{giddings}.  Even with the suppression factor there should be plenty of hard leptons or photons that give a clear indication of the decay of some non-perturbative state.

Finally we would like to add that even if the string scale turns out to be much larger than 1 TeV the results in this letter should be valid in any situation where there is a weakly coupled string theory and a brane on which gauge fields live surrounded by a large bulk into which the gauge fields cannot go.  However, it is difficult to imagine what the observable implications would be if the string scale was above the limit of non-negligible cosmic ray fluxes, i.e. a few tens or hundreds of TeV.

\section*{Acknowledgments}
It is a pleasure to thank Sergei Troitsky for his help.

\end{document}